\documentclass[twocolumn,aps,prx,twoside,superscriptaddress,floatfix,nofootinbib]{revtex4-1}

\usepackage{dcolumn}
\usepackage{float}
\usepackage{bm}
\usepackage[T1]{fontenc}
\usepackage{amsmath}
\usepackage{graphicx} 
\usepackage{sidecap}
\usepackage{tabularx}
\usepackage{multirow}
\usepackage{rotating} 
\usepackage{makecell}
\usepackage{hyperref}
\usepackage{mathtools}
\usepackage[version=4]{mhchem}
\usepackage{ragged2e}
\usepackage{chemformula} \setchemformula{kroeger-vink} 
\hypersetup{colorlinks=true, linkcolor=blue, citecolor=blue, urlcolor=blue}
\usepackage{tikz}
\usepackage[normalem]{ulem}

\makeatletter
\renewcommand\@biblabel[1]{#1.}
\makeatother

\newcommand{\super}[1]{\ensuremath{^{\textrm{#1}}}}
\newcommand{\sub}[1]{\ensuremath{_{\textrm{#1}}}}

\begin{document}

\title{Oxygen vacancies in bulk and at neutral domain walls in hexagonal YMnO$_3$}
\author{Sandra H. Skj{\ae}rv\o}
\affiliation{NTNU Norwegian University of Science and Technology, Department of Materials Science and Engineering, NO-7491 Trondheim, Norway}
\author{Didrik R. Sm{\aa}br{\aa}ten}
\affiliation{NTNU Norwegian University of Science and Technology, Department of Materials Science and Engineering, NO-7491 Trondheim, Norway}
\author{Nicola A. Spaldin}
\affiliation{ETH, Materials Theory, Wolfgang Pauli Str.~27, CH-8093 Z\"{u}rich, Switzerland}
\author{Thomas Tybell}
\affiliation{NTNU Norwegian University of Science and Technology, Department of Electronic Systems, NO-7491 Trondheim, Norway}
\author{Sverre M. Selbach}
\email{selbach@ntnu.no}
\affiliation{NTNU Norwegian University of Science and Technology, Department of Materials Science and Engineering, NO-7491 Trondheim, Norway}

\begin{abstract}
We use density functional calculations to investigate the accommodation and migration of oxygen vacancies in bulk hexagonal YMnO$_3$, and to study interactions between neutral ferroelectric domain walls and oxygen vacancies. Our calculations show that oxygen vacancies in bulk YMnO$_3$ are more stable in the Mn-O layers than in the Y-O layers. Migration barriers of the planar oxygen vacancies are high compared to oxygen vacancies in perovskites, and to previously reported values for oxygen interstitials in h-YMnO$_3$. The calculated polarization decreases linearly with vacancy concentration, while the out-of-plane lattice parameter expands in agreement with previous experiments. In contrast with ferroelectric perovskites, oxygen vacancies are found to be more stable in bulk than at domain walls. The tendency of oxygen vacancies to segregate away from neutral domain walls is explained by unfavorable Y-O bond lengths caused by the local strain field at the domain walls.
\end{abstract}

\maketitle
\section{Introduction}
Ferroelectric domain walls (DW), which separate regions of different ferroelectric polarization, have been demonstrated to possess distinctly different electronic properties from their parent bulk compounds \cite{catalan_domain_2012,meier_functional_2015,seidel_conduction_2009}. Nanoscale ferroelectric domain walls are natural structural interfaces that can be created, moved and erased by applied electric fields. 
Recent discoveries of metallic magnetic domain walls in an insulating antiferromagnet \cite{Ma.Science.350.538} and the ability to switch the conductivity of a ferroelectric domain wall on and off by an electric field \cite{mundy_functional_2017} hold great promise for future developments of domain wall-based circuitry \cite{meier_anisotropic_2012,meier_functional_2015}.

Enhanced domain wall conductivity was first demonstrated in BiFeO$_3$ and attributed to local structural changes giving an electrostatic potential step and a local reduction of the band gap \cite{seidel_conduction_2009}. At charged domain walls, one or more components of the polarization vectors of two adjacent domains meet either head-to-head or tail-to-tail. The resulting electrostatic fields can be screened by mobile charge carriers such as electrons or holes, leading to enhanced conductivity \cite{Sluka.NatCommun.4.1808,meier_anisotropic_2012,Morozovska.PhysRevB.86.085315}. 
The field screening charge carriers can arise from charge compensation of intrinsic point defects such as oxygen vacancies. Enhanced domain wall conductivity has been attributed to oxygen vacancies in improper ferroelectric hexagonal manganites \cite{kim_domain_2016, chen_oxygen-vacancy_2012, zhang_oxygen-vacancy-induced_2010, du_manipulation_2013, du_domain_2011} and in conventional ferroelectrics materials such as PbTi$_{1-x}$Zr$_x$O$_3$ (PZT) \cite{Guyonnet.AdvMater.23.5377} and BiFeO$_3$ \cite{gaponenko_towards_2015,Seidel.PhysRevLett.105.197603}.

In the hexagonal manganites, h-RMnO$_3$, both charged and neutral walls have shown increased conductivity \cite{choi_insulating_2010, meier_anisotropic_2012, wu_conduction_2012, campbell_hall_2016, mundy_functional_2017,schaab_optimization_2016,Ruff.PhysRevLett.118.036803,wu_low-energy_2017}. While the electrostatic field can drive an accumulation of electrons or holes at charged walls, at neutral walls the local elastic field is the most plausible driving force for charge carrier accumulation. Oxygen vacancies were predicted to accumulate at neutral domain walls from atomistic simulations \cite{jiang_atomistic_2015}, and hence charge compensating electrons could increase the conductivity.

\begin{SCfigure*} 
\includegraphics[width=0.65\textwidth]{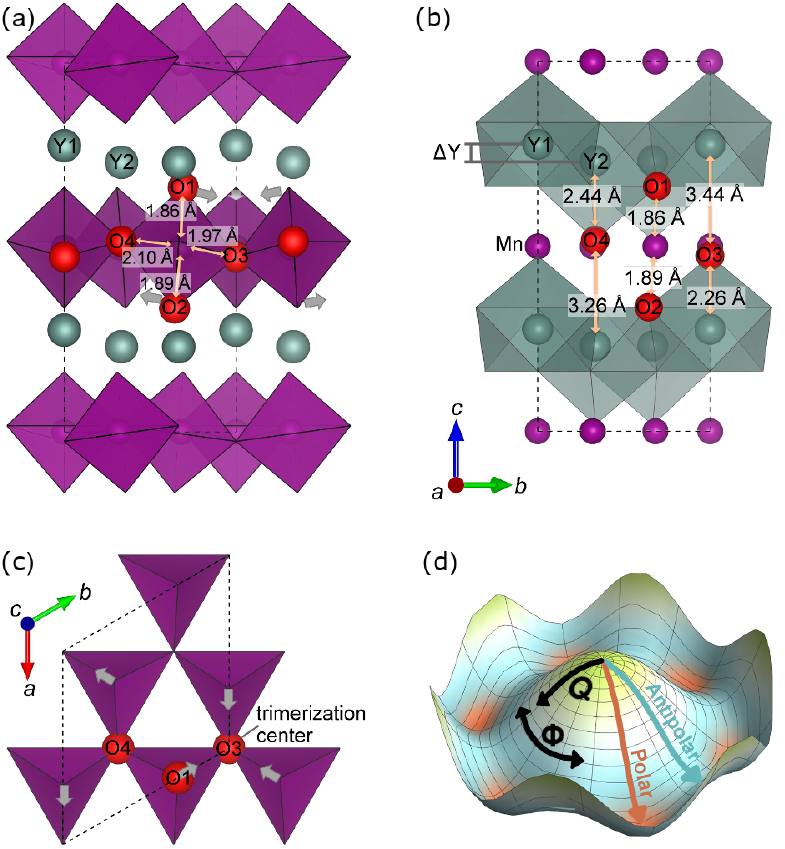}
\caption{(a) Unit cell of the YMnO$_3$ crystal structure visualized with the conventional MnO$_5$ trigonal bipyramidal polyhedra emphasizing the atomic Wyckoff sites (Y1-2, Mn, O1-4), up-down-down off-centering of the Y atoms and the tilting of the polyhedra. (b) Equivalent unit cell of YMnO$_3$ visualized with YO$_7$ polyhedra emphasizing the strong Y-O bonds. Out-of-plane bonds between the Y ions and the planar O3 and O4 ions in the Mn layers link the sublattices together. Selected bond lengths are given in the panels. (c) Trimerization tilting pattern of the MnO$_5$ polyhedra seen along the $c$ axis. We denote the O3 ion a trimerization center because the three surrounding MnO$_5$ polyhedra are tilting directly towards it. (d) Calculated Mexican hat energy landscape, following methods by Artyukhin \textit{et al.}~\cite{artyukhin_landau_2013}. $Q$ and $\Phi$ are the order parameter amplitude and angle, respectively. A high-symmetry $P6_3/mmc$ structure is found at the top, while the polar ground state with $P6_3cm$ symmetry is found in the six minima in the brim. The anti-polar structure with $P\bar{3}c1$ symmetry is found at the local maxima in the brim.}
\label{fig:structure}
\end{SCfigure*}

The YMnO$_3$ structure with space group $P6_3cm$ has layers of corner-sharing MnO$_5$ polyhedra separated by seven-coordinated Y, as illustrated for the prototypical YMnO$_3$ in Fig.~\ref{fig:structure}(a)-(b). In the Mn-O layers, groups of three MnO\sub{5} polyhedra tilt towards or away from the planar O3 sites, denoted trimerization centers, see Fig.~\ref{fig:structure}(c). The apical oxygen (O1 and O2) and planar oxygen (O3 and O4) sites in the MnO$_5$ polyhedra have different coordination environments in this layered structure. The MnO$_5$ tilting pattern (Fig.~\ref{fig:structure}(a),(c)) is strongly coupled to the alternating off-centering of Y along the $c$-direction, Fig.~\ref{fig:structure}b. These distortions are direct observables of the primary order parameter, the zone-boundary $K_3$ mode, here referred to as the \textit{trimerization} mode. Improper ferroelectric polarization arises from a shift of the Y layer with respect to the Mn-O layer, with polarization saturating at $\sim$6 $\mu$Ccm$^{-2}$ at room-temperature \cite{gibbs_high-temperature_2011, van_aken_hexagonal_2001,lilienblum_ferroelectricity_2015}. The coupling between the trimerization and polarization gives an energy landscape resembling a Mexican hat \cite{artyukhin_landau_2013}, Fig.~\ref{fig:structure}(d), with six minima corresponding to the combination of two polarization directions and three trimerization centers.
As the primary trimerization distortions of the Y-O and Mn-O sublattices are linked through out-of-plane Y-O bonds along the polar axis, we expect oxygen vacancies to perturb both the trimerization distortion and the ferroelectric polarization.

The hexagonal manganite structure can accommodate large concentrations of both oxygen vacancies \cite{overton_influence_2009, du_manipulation_2013} and interstitial oxygen ions \cite{skjaervo_interstitial_2016, remsen_synthesis_2011, abughayada_structural_2014}. Oxygen \textit{vacancy} formation is favored by high temperature and low $p$O$_2$ \cite{selbach_crystal_2012}, while enthalpy stabilized oxygen interstitials are favored at low temperatures and high $p$O$_2$ \cite{skjaervo_interstitial_2016}. The oxygen stoichiometry can thus be tuned through the thermal and atmospheric history of the materials. Under a finite vacancy concentration, vacancies are charge compensated by formal reduction of Mn in accordance to Eq.~\ref{eq:kroger}. 
\begin{flalign}
\ch{2 Mn_{Mn}^{x} + O_O^{x}} \ce{<=>} 2 \ch{Mn_{Mn}^{$\prime$} + V_O^{..}} + \sfrac{1}{2} \ch{ O_2~(g)}
\label{eq:kroger}
\end{flalign}

where \ch{V_O^{..}} and \ch{Mn_{Mn}'} denote an oxygen vacancy with relative charge +2 and a Mn$^{2+}$ cation, respectively. Oxygen vacancies can act as electron donors, giving enhanced conductivity at head-to-head domain walls, \cite{hassanpour_robustness_2016}, or reduced conductivity at tail-to-tail walls \cite{Holstad.PhysRevB.97.085143}.

Here we study the formation and migration of oxygen vacancies in YMnO$_3$ using density functional theory (DFT) calculations. We find a strong preference for formation of vacancies at the planar O4 sites in bulk, which are not trimerization centers, where they are elastically screened by the surrounding ions, and reduce the local polarization. Oxygen vacancies migrate most easily in the Mn-O layers between O4 sites. Close to neutral domain walls, the defect formation energy increases due to breaking of shorter Y-O bonds, suggesting that there is no driving force for accumulation of oxygen vacancies at these walls. Oxygen vacancies are thus \textit{not} likely to be the origin of local enhanced conductivity at neutral domain walls.

\section{Computational details} \label{sec:compdet}
Density functional calculations were performed using the VASP code \citep{kresse_efficient_1996, kresse_ultrasoft_1999} with the GGA PBEsol functional \citep{perdew_restoring_2008}. To better describe the Mn-$d$ on-site Coulomb interaction, we added a Hubbard $U$ of 5 eV within the spin-polarized GGA+$U$ implementation of Dudarev \citep{dudarev_electron-energy-loss_1998}, reproducing the experimental bandgap \citep{degenhardt_nonlinear_2001} and lattice parameters \citep{van_aken_hexagonal_2001}. The projector augmented wave method \citep{blochl_projector_1994} was used, treating Y(4s,4p,4d,5s), Mn(3s,3p,3d,4s) and O(2s,2p) as valence electrons, combined with a plane-wave cutoff energy of 550 eV. Brillouin zone integration was done on a $\Gamma$-centered 2$\times$2$\times$2 mesh for the 120 atom supercells and 1$\times$1$\times$1 mesh for the 3$\times$3$\times$2 supercells. The non-collinear magnetic structure was approximated by a collinear frustrated anti-ferromagnetic order (F-AFM) \citep{medvedeva_effect_2000}. 

The geometry of bulk structures was relaxed until the forces on the ions were below 0.01 eV/\AA~for 120 atom cells and 0.02 for 540 atom cells. Lattice parameters were kept fixed in all defect calculations, except for calculations of chemical expansion. Oxygen vacancy formation energies were calculated from\\ $E^f_{V_{\rm O}} = E^f_{\rm {YMnO_{3-\delta}}}-E^f_{\rm {YMnO_{3}}}+\mu_{\rm O}$, with the chemical potential of oxygen set to $\mu_{\rm O}$ = -4.5~eV. As YMnO$_3$ is prone to substantial oxygen non-stoichiometry, we consider neutral cells here, hence $E^f_{V_{\rm O}}$ is effectively the enthalpy of reduction. The ferroelectric polarization was calculated with a simple ionic point charge model using formal charges, as this method gives comparable values to Berry phase calculations for YMnO$_3$ \cite{kumagai_structural_2013}. Five images were used for the climbing image nudged elastic band (cNEB) \cite{henkelman_climbing_2000,henkelman_improved_2000} calculations of vacancy migration paths. 

Neutral domain wall cells were built following refs. \citep{kumagai_structural_2013,matsumoto_multivariate_2013}. 240 atom 8$\times$1$\times$1 supercells were used to ensure close-to-bulk properties within the domains. Magnetic order was continued across the domain walls for simplicity. For the domain wall calculations a k-point grid of 1$\times$4$\times$2 was chosen, and the geometry optimized until the forces on the ions were below 0.02 eV/\AA.
    
\begin{table}[b!]

\caption{\textbf{Relaxed bulk structure of YMnO$_3$}}
\justifying\noindent Relaxed atomic positions for YMnO$_3$ within the space group $P6_3cm$. Computational details are given in Section \ref{sec:compdet}. The relaxed lattice parameters are $a$ = 6.099 \AA~ and $c$ = 11.422 \AA.
\centering
	\begin{tabularx}{0.5\textwidth}{ X X X X }
&&&\\
	\hline
		Position & x & y & z \\ 
		\hline
		Y1 & 0 & 0 & 0.276 \\
		Y2 & 1/3 & 2/3 & 0.231 \\
		Mn & 0.334 & 0 & 0 \\
		O1 & 0.306 & 0 & 0.164 \\
		O2 & 0.640 & 0 & 0.337 \\
		O3 & 0 & 0 & 0.476 \\
		O4 & 1/3 & 2/3 & 0.021\\
		\hline
	\end{tabularx}
    \label{tab:bulk}
\end{table}

\subsection{Bulk crystal structure}
Before addressing structural accommodation of oxygen vacancies in bulk YMnO$_3$, we first calculated the ground state crystal structure using the computational details given in Section \ref{sec:compdet}. Our relaxed structural coordinates, presented in Table \ref{tab:bulk}, agree well with the experimental structure reported by Gibbs \textit{et al.} \cite{gibbs_high-temperature_2011}.

\subsection{Vacancy energetics}
\begin{figure} 
\centering
\includegraphics[width=0.45\textwidth]{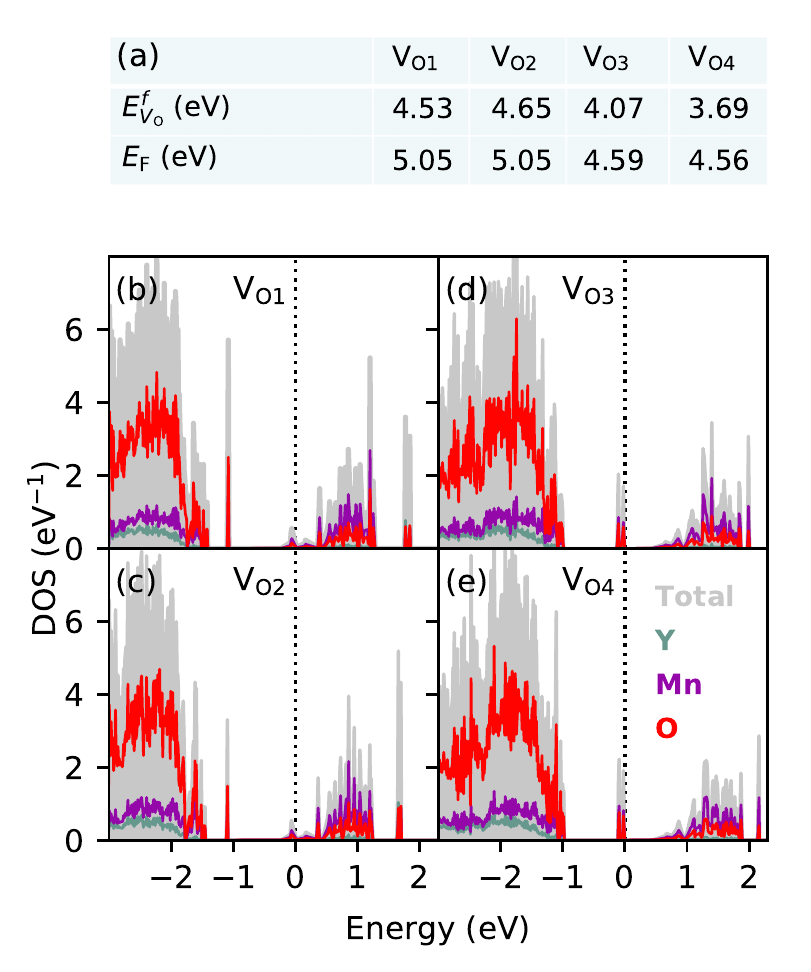}
\caption{(a) formation energies $E^f_{V_{\rm O}}$ and Fermi energies $E_F$ for all oxygen vacancy types calculated with neutral cells (2$\times$2$\times$1 supercells). (b)-(e) Electronic densities of states for the same cells, with the Fermi energy set to 0 eV.}
\label{fig:energy}
\end{figure}

Relaxation of structures with oxygen vacancies at the four different atomic sites show that the planar oxygen vacancies have significantly lower defect formation energy than apical oxygen vacancies, Fig.~\ref{fig:energy}(a). A simple chemical argument can be used to understand this difference: the local bonding environment of apical oxygen ions is more dominated by Y than by Mn. Conversely, the local coordination environment of planar oxygen ions is more dominated by their short bonds to Mn in the same layer. Since Y has a stronger affinity for oxygen and is more electropositive than Mn, the Y-O bonds are stronger than the Mn-O bonds, making the latter easier to break when creating a vacancy. This preference for planar oxygen vacancies over apical vacancies is not found in h-InMnO$_3$, consistent with In and Mn having similar electronegativities and oxygen affinities \cite{griffin_defect_2017}.

The lower formation energy for the planar vacancies compared to the apical vacancies is also evident from the electronic structures. The electronic density of states for each of the four vacant oxygen sites are presented in Fig.~\ref{fig:energy}(b)-(e). The electronic structures show that an oxygen vacancy reduces Mn and lifts the Fermi level above the valence band maximum. For the apical oxygen vacancies V\sub{O1} and V\sub{O2}, the Fermi level is lifted into the conduction band, with a Fermi energy of $\sim$5.0 eV. The charge compensating electrons are localized in Mn $d_{z^{2}}$ states, formally reducing two Mn\super{3+} to Mn\super{2+} (Eq.~\ref{eq:kroger}), with Bader charges of 1.72 and 1.47, respectively. For the planar oxygen vacancies V\sub{O3} and V\sub{O4}, the $d_{z^{2}}$ defect state at the Fermi level is only lifted into the band gap, with a Fermi energy of $\sim$4.5 eV. The charge compensating electrons are more localized for planar than apical vacancies, with Bader charges of 1.72 and 1.38 for Mn\super{3+} and Mn\super{2+}, respectively.

\begin{figure*} 
\centering
\includegraphics[width=\textwidth]{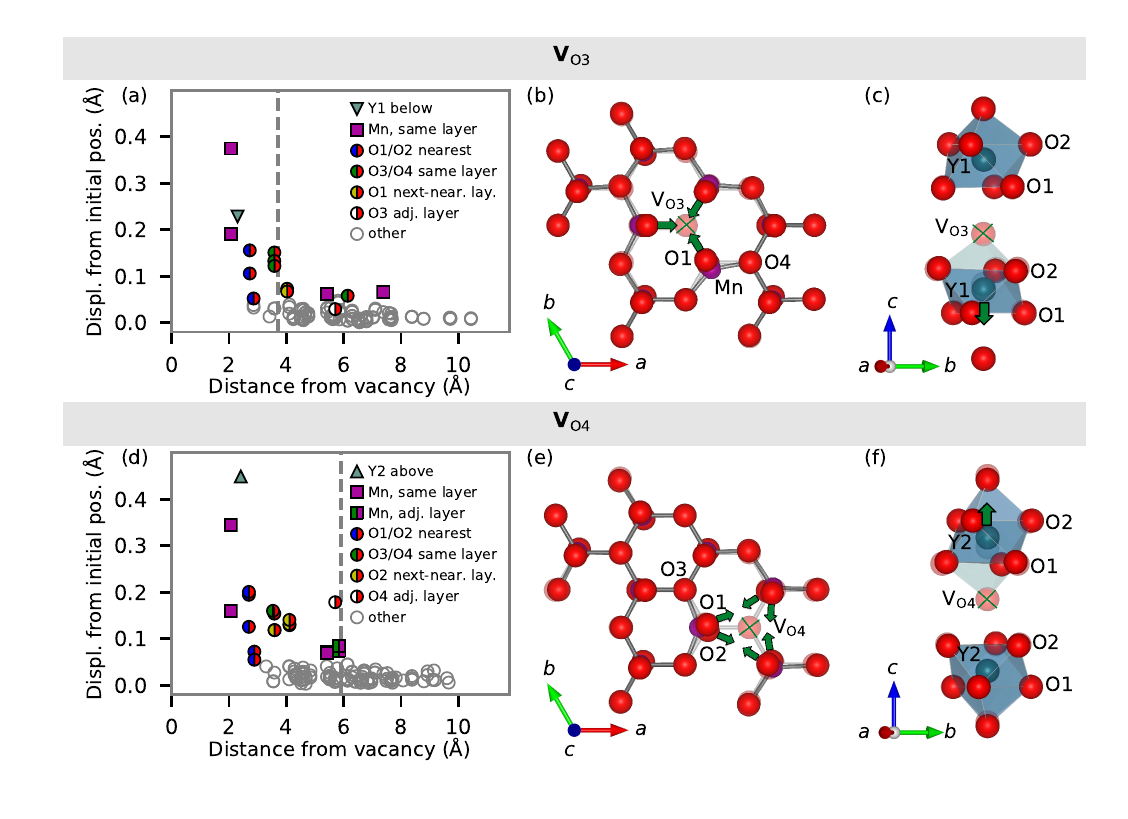}
\caption{Displacement from initial position of all atoms in 520 atom 3$\times$3$\times$2 cells as a function of distance from an O3 (a) or O4 (d) vacancy. Structural screening lengths are indicated by dashed vertical lines.
Structural relaxations around an O3 (b)-(c) or O4 (e)-(f) vacancy with the initial structure faded in the background. Green arrows indicate significant changes in bipyramidal tilting or Y position after relaxation.}
\label{fig:displacements}
\end{figure*}

The electronic density of states cannot explain the difference in formation energies between the two planar sites, favoring the O4 site by 0.37 eV. This indicates that the chemical bonding environments around the O3 and O4 vacancies are different; the O4 vacancies are stabilized further by elastic effects, which is discussed next.

\subsection{Structural accommodation of vacancies}
The difference in formation energy between O3 and O4 lies in their bonding to nearby Y atoms and the structural flexibility of the Y-O$_7$ polyhedra. We remind the reader that O3 is a trimerization center (Fig.~\ref{fig:structure}(c)). A combination of the trimerization distortion and the polar shift along $c$, results in shorter Y1-O3 bonds than Y2-O4 bonds (Fig.~\ref{fig:structure}(b)). Hence, breaking Y1-O3 bonds is expected to cost more energy than breaking Y2-O4 bonds.

The difference in formation energy between V\sub{O3} and V\sub{O4} is further rationalized by considering the local structural distortions. We quantify the distortions by defining a \textit{structural screening length} as the distance away from the defect where the perturbations of atoms are smaller than 0.1 \AA. Introducing planar oxygen vacancies causes only short range structural perturbations, as shown in Fig.~\ref{fig:displacements}(a)-(b). Significant perturbations are found in the Mn-O layer in which the vacancies are located, but the perturbations are effectively screened from the third coordination shell from the defect. The structural screening length of a V\sub{O3} is $\sim$4 \AA, while it is $\sim$6 \AA~for V\sub{O4}. The longer screening length for V\sub{O4} is mainly caused by a displacement of O4 in the adjacent Mn-O layer. Perturbation out-of-plane is also seen for the closest Y1 and Y2, directly above or below the vacancies. 

It may appear counterintuitive that the larger spatial extent of distortions around the O4 vacancy compared to the O3 vacancy still leads to a lower formation energy. To rationalize this, the different structural distortions caused by the two planar vacancies are analyzed with respect to the energy landscape of the order parameter $Q, \Phi$ of the trimerization (Fig.~\ref{fig:structure}(d)). 
A V\sub{O3} is accommodated by an increased apical tilt angle, $\alpha$\sub{A}, of the three closest trigonal bipyramids (Fig.~\ref{fig:displacements}(c), and a subtle repulsion of the closest Y1 (Fig.~\ref{fig:displacements}(d)). These perturbations correspond to an increase in the trimerization amplitude $Q$ in the Mexican hat energy landscape (Fig.~\ref{fig:structure}(d)), which comes with a high cost and a short structural screening length.
The more stable V\sub{O4}, on the other hand, is mainly accommodated by rotation of the closest trigonal bipyramids (Fig.~\ref{fig:displacements}(e)), and a significant repulsion of the closest Y2 (Fig.~\ref{fig:displacements}(f)). These perturbations come with a lower cost, as they correspond to changing the trimerization angle $\Phi$ in the brim of the Mexican hat energy landscape. This link between structural distortions and the order parameter energy landscape \cite{skjaervo_unconventional_2017} explains the strong driving force for vacancies favoring the O4 site in bulk YMnO$_3$.

It is worth mentioning, that 0K first principles calculations predict a larger trimerization amplitude than that observed at finite temperatures. A smaller trimerization amplitude at finite temperatures is expected to reduce the difference in defect formation energy between V\sub{O3} and V\sub{O4}.

\subsection{Effect on polarization and lattice}
\begin{figure}
\centering
\includegraphics[width=0.5\textwidth]{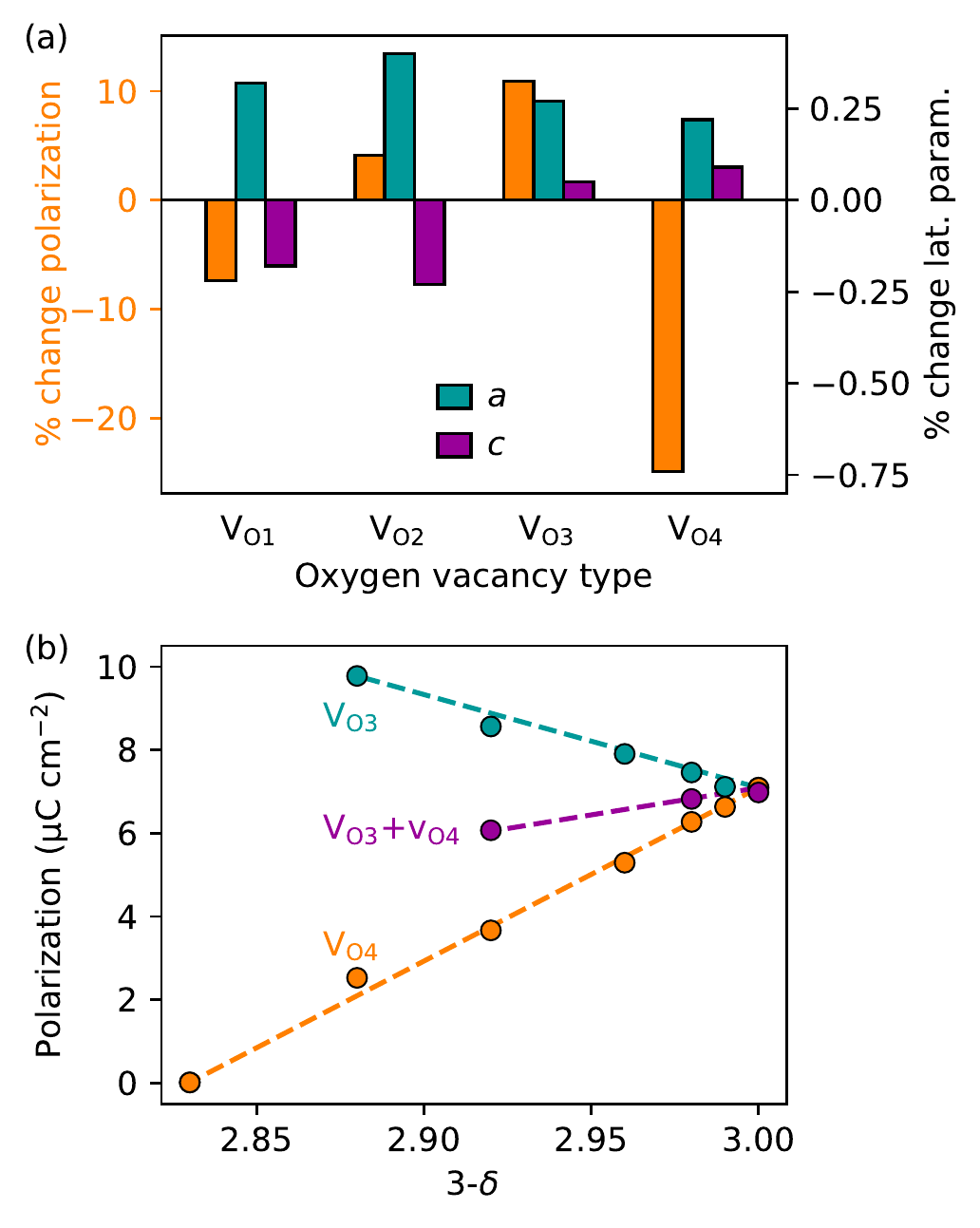}
\caption{(a) \% change in polarization and lattice parameters $a$ and $c$ of 120 atom supercells with one oxygen vacancy relative to stoichiometric cell, where both atomic positions and cell volume have been allowed to relax. Calculations with relaxation of only atomic positions gave similar results for the polarization. (b) Polarization as a function of concentration of V$_{\rm O3}$, V$_{\rm O4}$ or both.}
\label{fig:polarization}
\end{figure}

The out-of-plane structural perturbations of Y naturally affects the local polarization. The calculated polarization for each of the four vacancies shows no correlation with the $c$ lattice parameter, as shown in Fig.~\ref{fig:polarization}(a). This is expected for an improper ferroelectric in which the polarization does not directly couple to strain, and thereby to the lattice parameters. 

The lattice expands within the \textit{ab} plane for all four vacancy positions, but it expands the least for the most stable V$_{\rm O4}$ vacancy, as shown in Fig.~\ref{fig:polarization}(a). For V$_{\rm O4}$ in the 2$\times$2$\times$1 supercell (YMnO$_{2.96}$), the expansion of the $ab$ plane is equal to 0.22\%, in good agreement with experimental XRD results of 0.30\% \cite{overton_influence_2009}.
The lattice expands by 0.09\% along $c$ for the V$_{\rm O4}$ vacancy in the 2$\times$2$\times$1 supercell (YMnO$_{2.96}$). This is in qualitative agreement with experimental XRD results of 0.42\% expansion for a similar oxygen vacancy concentration in YMnO$_{2.95}$ \cite{overton_influence_2009}. The quantitative difference may stem from the anticipated Boltzmann distribution of vacancies at ambient temperatures, compared to the periodic image ordering at 0K inherent in our DFT calculations. 

The short structural screening length around the vacancies allows for large vacancy concentration without significant defect-defect interactions. Thus, the polarization changes linearly as a function of vacancy concentration, as shown in Fig.~\ref{fig:polarization}(c). The polarization in improper ferroelectric YMnO$_3$ is robust with respect to the most stable V$_{\rm O3}$ vacancies, and becomes zero only at a high concentration corresponding to a stoichiometry of YMnO$_{2.833}$, where one V$_{\rm O4}$ is present in a 30 atom cell. The polarization increases with increasing concentration of the less stable V$_{\rm O3}$ until a stoichiometry of YMnO$_{2.875}$ is reached, before dropping abruptly for YMnO$_{2.833}$.

\subsection{Oxygen vacancy migration}
\begin{figure}
\includegraphics[width=0.5\textwidth]{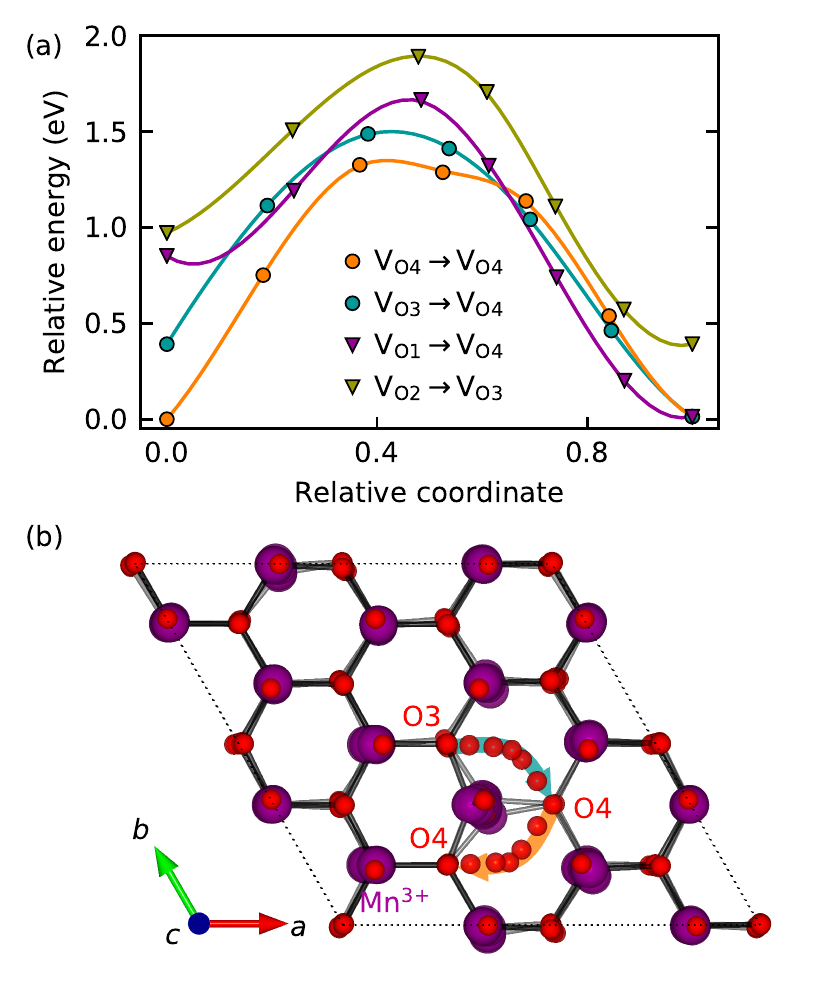}
\caption{(a) In-plane (circles) and out-of-plane (triangles) migration barriers for an oxygen vacancy in bulk YMnO$_3$. In-plane migration: The vacancy can migrate between O4 and O3 sites, or between two O4 sites. Out-of-plane migration: the vacancy can migrate between O1 and O4 sites or between O2 and O3 sites. (b) Corresponding calculated in-plane migration paths. Arrows with colors corresponding to plots in panel (a) show the migration paths of lattice oxygen to vacant oxygen sites. Multiple positions of the ions correspond to images in the cNEB calculations and illustrate the structural breathing during vacancy migration.}
\label{fig:migration}
\end{figure}

Oxygen ion migration energy barriers give clues to how quickly the stoichiometry is expected to equilibrate with the surroundings, and in layered YMnO$_3$ they reveal the preferred lattice planes for ionic transport. Calculated paths for migration between planar oxygen sites and between apical and planar oxygen sites are presented in Fig.~\ref{fig:migration}. The migration path between two O4 sites has the lowest energy barrier of $\sim$1.3 eV, while the other in-plane path between an O3 and an O4 site has a barrier of $\sim$1.5 eV. Out-of-plane migration is less favorable, with barriers of $\sim$1.7 eV and $\sim$1.9 eV for the O1 to O4 and O2 to O3 paths, respectively. Experimental reports have shown that oxygen vacancies move in plane under an applied electron beam \cite{zhang_oxygen_2015}. Our calculated oxygen vacancy migration barriers are high compared to state-of-the-art oxygen conducting perovskites \cite{chroneos_oxygen_2010}; in Ba$_{0.5}$Sr$_{0.5}$Co$_{0.8}$Fe$_{0.2}$O$_{3-\delta}$ for example the migration barrier is $\sim$0.4-0.5 eV \cite{Shao.Nature.431.170,Kotomin.SolidStateIon.188.1}. The vacancy migration barrier in YMnO$_3$ is also much higher than that for interstitial oxygen in h-YMnO$_3$ \cite{skjaervo_interstitial_2016}, hence we predict that oxygen transport in hexagonal manganites occurs predominantly in the Mn-O planes, and is strongly favored in oxygen rich compared to oxygen poor materials. 

\section{Vacancies at neutral domain walls}

\begin{figure}[b!]
\includegraphics[width=0.53\textwidth]{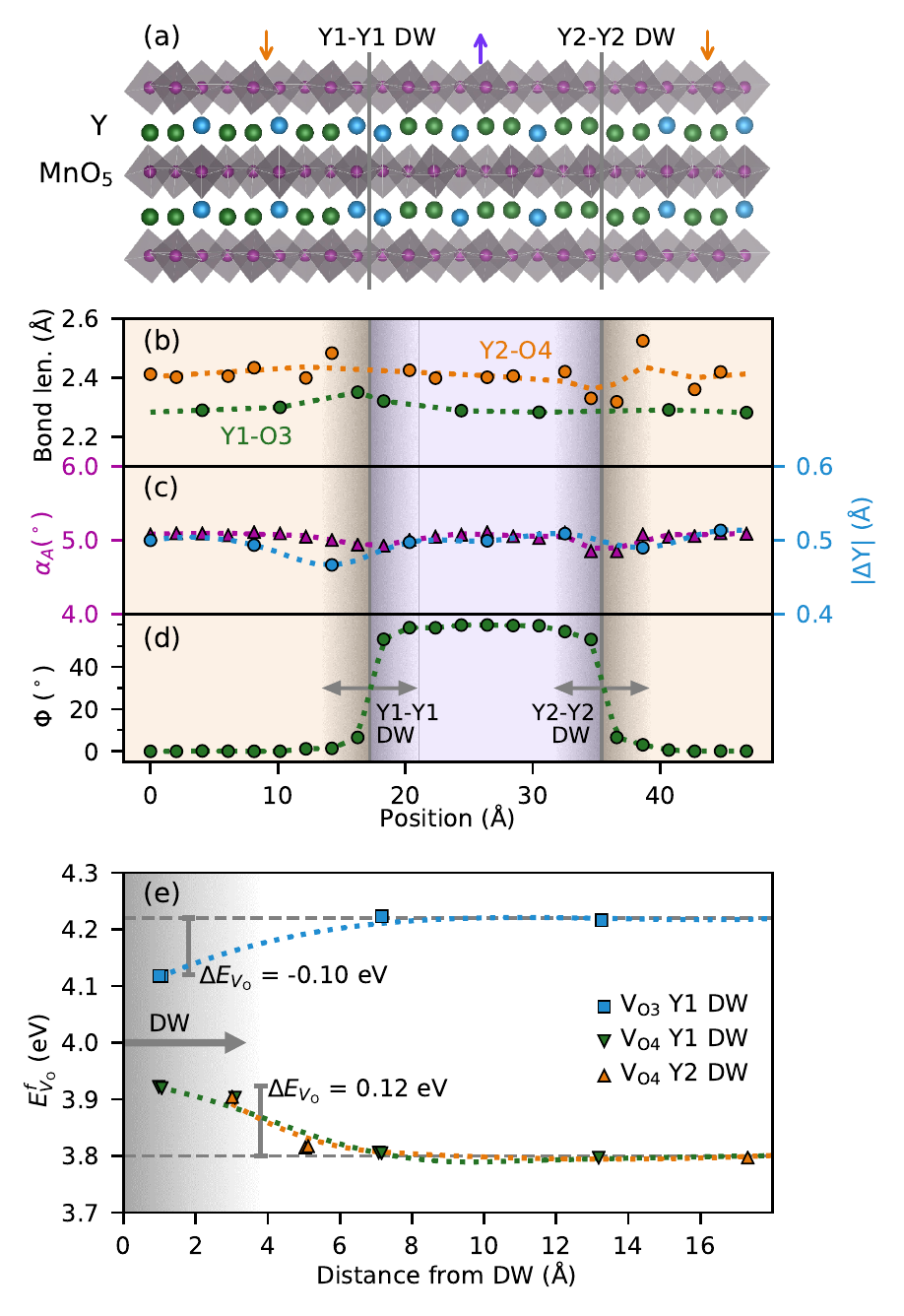}
\caption{(a) Stoichiometric 8$\times$1$\times$1 supercell with two domains separated by (210) neutral Y1-Y1 and Y2-Y2 terminated domain walls. Orange and purple arrows denote the direction of the polarization. Y1 and Y2 are colored blue and green, respectively. (b) Out-of-plane Y1-O3 and Y2-O4 bond lengths across the domain wall cell. (c) Observables of the trimerization order parameter amplitude $Q$: Apical tilt angle, $\alpha_A$ of the MnO$_5$ bipyramids, and the Y corrugation $\Delta$Y = $c(z_{\rm Y1}- z_{\rm Y2})$. (d) Trimerization order parameter angle $\Phi$ calculated from the tilting direction of the trigonal bipyramids. (e) Defect formation energy $E^f_{V_{\rm O}}$ for O3 and O4 vacancies as a function of distance to Y1 or Y2 terminated domain walls. The oxygen chemical potential is set to -4.5 eV.}
\label{fig:DWcell}
\end{figure}

Having established the energetics and structural effects of oxygen vacancies in bulk YMnO$_3$, we now turn our attention to oxygen vacancies at neutral ferroelectric domain walls. At neutral domain walls, it follows from symmetry considerations that the domains are terminated by either Y1 or Y2 planes\cite{kumagai_structural_2013,choi_insulating_2010}. The structural relaxation of domain walls, presented in Fig.~\ref{fig:DWcell}(a), shows that the walls resemble structural "stacking faults", in line with previous TEM images from Matsumoto \textit{et al.} \cite{matsumoto_multivariate_2013} and DFT studies from Kumagai \textit{et al.} \cite{kumagai_structural_2013}. None of the Y atoms relax to the high-symmetry positions which would give the antipolar $P\bar{3}c1$ symmetry found as a special point in the local maxima of the brim of the Mexican hat energy landscape (Fig.~\ref{fig:structure}(d). The gradual distortions across a domain wall lower the symmetry to $P3c1$, which is intermediate between polar $P6_3cm$ and antipolar $P\bar{3}c1$, and a subgroup of both. As the O3 and O4 ions are situated directly below or above an Y1 or Y2, respectively, their distances to the Y1 or Y2 terminated domain walls vary accordingly, Fig.~\ref{fig:DWcell}(b). The trimerization amplitude $Q$ is subtly reduced at the domain walls, as seen from $\alpha$\sub{A} and $\Delta$Y as a function of cell position, Fig.~\ref{fig:DWcell}(c). The gradual 60$^{\circ}$ change of $\Phi$ over a $\sim$6 \AA~range across the walls, denotes the domain wall width, marked by the gray arrows in Fig.~\ref{fig:DWcell}(d). The variation of the order parameter across the domain walls, and the corresponding local lattice strain, are anticipated to affect the stability of oxygen vacancies.

To study the interaction between the planar oxygen vacancies and the neutral domain walls, the defect formation energies in 8$\times$1$\times$1 supercells were calculated for single oxygen vacancies at varying distances from the two types of neutral domain walls. To quantify the driving force for accumulation of oxygen vacancies at neutral domain walls, the segregation enthalpy was calculated as $\Delta E^f_{V_{\rm O}} = E^f_{\textrm{DW}} - E^f_{\textrm{bulk}}$. As shown in Fig.~\ref{fig:DWcell}(e), V\sub{O3} is attracted to the domain walls with a segregation enthalpy of $\Delta E^f_{V_{\rm O}}$ = -0.10 eV. In contrast, V\sub{O4} is repelled by the domain walls with a segregation enthalpy of $\Delta E^f_{V_{\rm O}}$ = 0.12 eV. We note that the formation energies of the vacancies only deviate from bulk values when the vacancies are within a distance from the domain wall approximately equal to the structural screening lengths found in bulk materials, Fig.~\ref{fig:displacements}(a),(d).

To assess the effect of cell size in the $ab$ plane, calculations with 4$\times$2$\times$1 supercells were performed. These calculations confirmed the tendency of oxygen vacancies to segregate away from neutral domain walls. The doubling of the unit cell along the $b$ axis reduces the vacancy-vacancy repulsion between periodic images, but gives qualitatively similar results for the segregation enthalpies with $\Delta E^f_{V_{\rm O}}$ = -0.14 eV for V\sub{O3} and $\Delta E^f_{V_{\rm O}}$ = 0.07 eV for V\sub{O4}. Oxygen vacancies thus prefer to be at O4 sites in bulk rather than at neutral domain walls. The opposite tendencies of accumulation found for V\sub{O3} and V\sub{O4} can be rationalized from the Y-O bond lengths (Fig.~\ref{fig:DWcell}(b)), keeping in mind the greater affinity of Y than Mn for oxygen. The Y1-O3 bonds are longer at the Y1-Y1 domain walls (2.33 \AA) than in bulk (2.30 \AA), indicating that breaking this bond at the domain wall will cost less energy than in bulk. Oppositely, Y2-O4 bonds are shorter at the Y2-Y2 domain walls (2.36 \AA) than in bulk (2.41 \AA), hence it costs less energy to break these bonds in bulk than at neutral domain walls.

\section{Summary and discussion}
In summary we find that oxygen vacancies are more stable in the Mn-O layers than at apical sites, where they are easily charge compensated by formally reducing two of the closest Mn\super{3+} to Mn\super{2+}. The electropositive character of Y and the flexibility of the YMnO$_3$ structure further favor vacancies at the planar O4 sites in bulk, minimizing the cost of breaking Y-O bonds. The higher cost of vacancies on O3 sites is explained by the shorter bond length to Y, and the energy cost of resulting perturbations of the trimerization amplitude $Q$. Vacancies on O4 sites in bulk distort mainly the trimerization angle $\Phi$ and reduce the local polarization, as breaking the Y-O\sub{P} bond reduces the coupling between Y and Mn sublattices. 
Migration of oxygen vacancies is found to occur preferentially in the Mn-O layer, with a high migration barrier compared to oxygen interstitials, and vacancies in perovskite oxides.
The vacancies are not predicted to segregate to neutral 180$^{\circ}$ domain walls. This is because of the lower cost of breaking Y2-O4 bonds in bulk compared to at domain walls, and the associated local strain field at domain walls affects the Y-O bond lengths. Oxygen vacancies are hence not likely to be the origin of observed enhanced conductivity at neutral domain walls in hexagonal manganites.

A preference for planar oxygen vacancy formation has been indicated experimentally by X-ray diffraction \cite{overton_influence_2009} and transmission electron microscopy \cite{cheng_electronic_2016,zhang_oxygen_2015}. 
However, distinguishing between the two planar oxygen vacancies in experiments is challenging. Although symmetry inequivalent, the two planar sites have previously been assumed to have equal occupancy of vacancies \cite{overton_influence_2009}, but we believe this is only true for high concentrations where the overall amplitude of the order parameter of trimerization is strongly reduced. Previous L(S)DA+U DFT studies \cite{cheng_electronic_2016,cheng_manipulation_2016} and atomistic simulation studies \cite{jiang_atomistic_2015} have shown preference for vacancies at the O4 site, in agreement with our results, although reporting a smaller energy difference between V\sub{O3} and V\sub{O4}. This quantitative difference may stem from the different lattice parameters, and concomitantly the geometry of the YO$_7$ polyhedra, obtained from L(S)DA and PBEsol calculations.

The positive domain wall segregation enthalpy for the most stable V\sub{O4} from our calculations contradicts previous atomistic simulations \cite{jiang_atomistic_2015}. These qualitative differences may stem from the fact that atomistic simulations with classical potentials can not adequately consider anisotropic electron density distributions, and thereby do not capture all covalent contributions to the chemical bonding. 

It has been shown that large concentrations of oxygen vacancies reduce both the trimerization amplitude and the ferroelectric polarization \cite{overton_influence_2009}, and that introduction of vacancies can enhance the mobility of the ferroelectric domain walls \cite{du_manipulation_2013}. Previous studies further show that the domain structure evolves from the characteristic clover-leaf patterns into stripe-like patterns in highly reduced samples \cite{chen_oxygen-vacancy_2012}, and that the domain walls become more mobile under electric field poling \cite{zhang_oxygen_2015}. This observed higher domain wall mobility can be explained by the reduced trimerization distortions when high concentrations of vacancies are introduced. Furthermore, the strong preference for vacancies at O4 sites, and their repulsion from the neutral domain walls may further favor stripe-like domains.

Although the elastic field at neutral domain walls repels oxygen vacancies, the situation can be very different at charged domain walls, where both elastic and internal electric fields will affect the stability of oxygen vacancies. Electric fields at charged head-to-head and tail-to-tail walls are intuitively expected to favor depletion and accumulation, respectively, of point defects with a positive relative charge such as oxygen vacancies. The possible competition and cooperation between elastic and electric fields for segregation of oxygen vacancies at charged domain walls in improper ferroelectrics, will be the topic of future studies.

In conventional ferroelectric perovskites, accumulation of oxygen vacancies at ferroelectric domain walls has been well-known for decades \cite{Postnikov.JPhysChemSolids.31.1785,Warren.ApplPhysLett.67.1426}. Oxygen vacancies are known to pin domain wall movement, causing hardening of PZT \cite{Morozov.JApplPhys.107.034106,Marsilius.JAmCeramSoc.93.2850,MorozovJApplPhys.118.164104} and fatigue in ferroelectric memories \cite{aggarwal_point_1998,dawber_model_2000}. Both seminal and recent DFT studies of oxygen vacancies at 180$^{\circ}$ and 90$^{\circ}$ domain walls in PbTiO$_3$ also confirm the tendency of oxygen vacancy accumulation and pinning in ferroelectric perovskites \cite{He.PhysRevB.68.134103,Chandrasekaran.PhysRevB.93.144102,Chandrasekaran.PhysRevB.88.214116}. 

While we do not find a tendency for oxygen vacancies to accumulate at neutral domain walls in YMnO$_3$, the situation is opposite for oxygen $interstitials$ \cite{schaab_electrical_2018}. Recent findings that the cation stoichiometry can also differ at domain walls compared to bulk materials \cite{farokhipoor_artificial_2014,Rojac.NatMater.16.322} further highlight the potential for actively engineering domain wall properties by changing the chemical composition relative to bulk materials. 

We hope our analysis of why oxygen vacancies will not segregate to neutral domain walls in YMnO$_3$ -- \textit{chemical bonds} and \textit{order parameter perturbations} -- will inspire efforts towards a generic understanding of the factors governing point defect-domain wall interactions in ferroelectrics and ferroelastics.


\bibliography{ReferenceLibrary}

\end{document}